\documentclass[runningheads,a4paper]{llncs}
\usepackage{amssymb}
\usepackage{graphicx}
\usepackage{booktabs}
\usepackage{listings}
\usepackage{xcolor}
\usepackage{hyperref}
\usepackage{caption}
\usepackage{subcaption} 
\usepackage{url}
\usepackage{color}
\usepackage{subfiles}
\newcommand{\orcid}[1]{\href{https://orcid.org/#1}{\includegraphics[scale=0.02]{figures/ORCIDiD_icon128x128.jpg}}} 
\usepackage{subfiles}
\usepackage{multirow}
\usepackage{cite}
\usepackage[nomargin,inline,marginclue,draft]{fixme}
\fxsetup{status=draft}
\fxsetup{theme=color, mode=multiuser} \FXRegisterAuthor{me}{ame}{\color{red}Me}

\hypersetup{
    colorlinks=true,
    citecolor=blue,
    urlcolor=purple,
}

\begin{document}
\mainmatter


\title{Detecting Histologic \& Clinical Glioblastoma Patterns of Prognostic Relevance}

\titlerunning{Glioblastoma Patterns of Prognostic Relevance}
\author{
    Bhakti Baheti\inst{1,2,3}
    \and
    Sunny Rai\inst{4}
    \and
    Shubham Innani\inst{1,2,3}
    \and
    Garv Mehdiratta\inst{1}
    \and
    Sharath Chandra Guntuku\inst{4,5}
    \and
    MacLean P. Nasrallah \inst{1,2}
    \and
    Spyridon Bakas\inst{1,2,3,*}
    }

    \institute{Center For Artificial Intelligence and Data Science for Integrated Diagnostics (AI\textsuperscript{2}D) and Center for Biomedical Image Computing and Analytics (CBICA), University of Pennsylvania, Philadelphia, PA, USA
    \and
    Department of Pathology and Laboratory Medicine, Perelman School of Medicine, University of Pennsylvania, Philadelphia, PA, USA
    \and
    Department of Radiology, Perelman School of Medicine, University of Pennsylvania, Philadelphia, PA, USA
    \and
    Department of Computer and Information Science, School of Engineering and Applied Science, University of Pennsylvania, Philadelphia, PA, USA
    \and
    Leonard Davis Institute of Health Economics, University of Pennsylvania, Philadelphia, PA, USA
    \\
    \textbf{*} Corresponding author\\
    \email{\{sbakas@upenn.edu\}}}
\authorrunning{Baheti et al.}

\maketitle
\begin{abstract}
Glioblastoma is the most common and aggressive malignant adult tumor of the central nervous system, with a grim prognosis and heterogeneous morphologic and molecular profiles. Since adopting the current standard-of-care treatment 18 years ago, no substantial prognostic improvement has been noticed. Accurate prediction of patient overall survival (OS) from histopathology whole slide images (WSI) integrated with clinical data using advanced computational methods could optimize clinical decision-making and patient management. Here, we focus on identifying prognostically relevant glioblastoma characteristics from H\&E stained WSI \& clinical data relating to OS. The exact approach for WSI capitalizes on the comprehensive curation of apparent artifactual content and an interpretability mechanism via a weakly supervised attention-based multiple-instance learning algorithm that further utilizes clustering to constrain the search space. The automatically placed patterns of high diagnostic value classify each WSI as representative of short- or long-survivors. Further assessment of the prognostic relevance of the associated clinical patient data is performed both in isolation and in an integrated manner, using XGBoost and SHapley Additive exPlanations (SHAP). Identifying tumor morphological \& clinical patterns associated with short and long OS will enable the clinical neuropathologist to provide additional relevant prognostic information to the treating team and suggest avenues of biological investigation for understanding and potentially treating glioblastoma.  
\end{abstract}


\section{Introduction}
\label{section:introduction}

Glioblastoma (GBM) is the most common malignant adult brain tumor with heterogeneous morphological and molecular profiles coupled with grim prognosis (median overall survival=14 months) \cite{brennan2013somatic, sottoriva2013intratumor, verhaak2010integrated}. Evaluation of GBM glass slide tissue sections and clinical data has always been the routine front-line assessment to provide disease diagnosis and essential information for treatment and management. Even though this front-line assessment requires expert pathologists, 
it remains the gold standard. Recent technological advancements have led to the digital pathology era as conventional glass slides are more commonly scanned into high-quality digitized tissue sections - also known as whole slide images (WSI). These digitization efforts have further led to significant progress in the proliferation of computational analysis methods for histopathology imaging, particularly for gaining novel insights from population-based studies.

Prognostic stratification of GBM patients from radiology imaging has been extensively explored \cite{bakas2020overall, rathore2018deriving,macyszyn2015imaging,beig2021sexually}, but with a notable limit on the performance of their prognostic predictions that could be potentially addressed by computational integrative analysis of histopathology and clinical data. Prognostic stratification using WSI is a challenging task due to several reasons, including: i) the large size of WSI ($\sim$ 100,000 pixels), ii) the variations of biological structures representative of the tumor heterogeneity, and iii) availability of prognostic labels at the patient level, while each patient may have multiple associated WSI. The literature on computational analysis of histopathology can be divided into approaches analyzing the entire WSI and those focusing on limited regions of interest (ROI). Since WSI handling is challenging and resource-demanding, ROI-based methods have been preferred in the field, specifically the analysis of specific annotated ROIs marked by pathologists \cite{barker2016automated, cheng2018identification, mobadersany2018predicting, zhu2016deep}. Furthermore, with the increase in computational power combined with broader availability, numerous supervised and unsupervised approaches continue to be developed for WSI classification \cite{zhu2017wsisa, yao2020whole, kather2019predicting}.
Regarding the prediction of OS from WSI, many approaches have been developed for multiple cancer types\cite{yao2016imaging, zhu2016deep, baheti2022epco}. 
However, frequently current approaches rely predominantly on hand-crafted imaging features extracted from a small set of manually-labeled WSI patches that render these methods unable to capture the heterogeneous morphology of GBM \cite{zhu2017wsisa}. In addition, interrogating and understanding features that drive decisions for determining short and long OS predictions could reveal avenues for improvement in human expert assessments.

Beyond imaging features, and mimicking the diagnostic process of the neuropathologist, assessing patient clinical features is crucial in identifying OS patterns. Various approaches focusing on clinical data are available in the literature, mainly targeting data integration to predict patient OS\cite{CHEN2022865}. Integrating data from multiple modalities, such as WSI and clinical data, can provide a more comprehensive understanding of a disease by combining information from different perspectives. WSI can provide detailed images of tissue samples that can be used to identify specific features at the microscopic level, such as cellular and nuclear atypia, proliferation, vascular features, necrosis, and infiltrative tumor cells. Clinical data, in contrast, provide high-level information about each patient, such as age, sex, medical history, and treatment outcomes. By interpreting these two data sources in tandem, better associations can be made on how morphological patterns (observed in WSI) and clinical features are associated with patient outcomes. This can lead to new insights into disease mechanisms and potential treatment strategies. While performance is an important consideration when accounting for data from multiple modalities, the interpretability of this data is critical for understanding the underlying biology of a disease. Using clinical data in tandem with WSI, researchers can improve the performance of OS prediction and gain a deeper understanding of the factors contributing to disease progression and treatment response.

Although almost all GBM have relatively aggressive behavior, there is variation in patient survival, and if individual tumor behavior could be better understood, the clinical team would have an advantage in better designing a specific treatment regimen to optimize survival and quality of life on an individual patient basis. With that motivation, this study focuses on the prognostic stratification of GBM patients between those with short and long post-surgical OS based on WSI imaging features and clinical data. To this end, we propose a comprehensive WSI curation to remove any known artifactual content, followed by a weakly-supervised attention-based multiple-instance-learning algorithm. The interpretability mechanism of the latter, by attention heatmaps, allows morphology pattern analysis from distinct histologic GBM sub-compartments of prognostic relevance. The benefit of integrating clinical data for predicting OS is evaluated in parallel. After merging both modalities, WSI patterns, and clinical data are combined through late fusion to examine the change in the overall performance of the prognostic stratification. 
Complex interactions between subgroups and patients' OS are further explored by investigating the patient OS in subgroups based on sex (male vs. female) and MGMT promoter methylation status (methylated vs. unmethylated). Overall, the proposed approach provides an interpretable, accurate way of predicting GBM patient prognosis, potentially improving treatment decisions and patient outcomes.
 
Quantitative and qualitative analysis coupled with performance evaluation is conducted on the publicly available TCGA-GBM \cite{tcgagbm} and TCGA-LGG \cite{tcgalgg} data collections, following their reclassification according to the latest WHO 2021 CNS tumor classification criteria \cite{louis20212021}. Section \ref{section:methods} describes the data used in our study and details of the proposed method. Results are presented in section \ref{section:results} followed by discussion in section \ref{section:discussion}.

\section{Methods}
\label{section:methods}
\subsection{Patient population}
    The publicly-available TCGA-GBM \cite{tcgagbm} and TCGA-LGG \cite{tcgalgg} data collections, available through The Cancer Imaging Archive (TCIA) \cite{clark2013cancer}, have been selected for the quantitative and qualitative performance evaluation and interpretation of the proposed work. Specifically, a board-certified neuropathologist (MPN), with 9 years of experience working on brain glioma, has reclassified these two collections according to the 2021 WHO classification of CNS tumors \cite{louis20212021} to identify all GBM IDH-wildtype (CNS WHO grade 4) cases. The TCGA-LGG collection includes histologically low-grade astrocytomas that (based on the 2021 WHO CNS classification criteria \cite{louis20212021}) are now understood to be GBM, given the presence of molecular features that specify the disease process and distinct tumor evolution. Thus, histologically low-grade astrocytomas from TCGA-LGG that have molecular features of GBM (also known as \textit{`molecular GBM'}) are identified and included in this study's cohort to create an algorithm that applies to all clinical GBM, as currently defined by the WHO. In a related manner, specific TCGA-GBM cases have been excluded from this study's cohort as their molecular profile does not render them GBM, according to the currently WHO-defined GBM entity. The collective data are divided into 80\%-10\%-10\% ratio for training, validation, and testing purposes, using Monte Carlo simulation \cite{xu2001monte}. 

    For the subgrouping analysis, the reclassified dataset consists of 188 cases, with 112 male and 76 female patients. Additionally, out of the 188 cases, MGMT promoter methylation status is available for 109 patients, with 59 and 50 patients having methylated and unmethylated status, respectively. We have divided the data into 10-fold Monte Carlo cross fold validation while training the model. We report the average performance of the model across 10-folds for Sex - Male, Female; MGMT - Methylated, Unmethylated for imaging, clinical and fusion modalities in Table \ref{tab:subgroup}. 

    \subsubsection{Imaging Data}
    \label{section:dataset}
    A single Hematoxylin \& Eosin (H\&E)-stained WSI from each available case of the reclassified TCGA-GBM and TCGA-LGG collections is utilized. Because frozen sections include hydration artifacts due to freezing, this study's analysis focuses only on permanently-fixated Formalin-Fixed Paraffin-Embedded (FFPE) slides. Although multiple (varying numbers of) WSIs exist per patient ranging from 1 to 15, we selected only one WSI per patient for uniformity. The included WSI was chosen by an expert neuropathologist (MPN) based on quality and tumor proportion. These cases are included irrespective of their maximum apparent magnification level (20X or 40X). 

    \subsubsection{Clinical Data}
    Clinical data comprises a variety of clinical/demographic and diagnostic molecular parameters collected during ongoing treatment. A total of $15$ features (Table \ref{tab:xgbFeatures}) were carefully selected from a large pool of clinical data provided on the TCGA portal. Of these 15 features, 3 are continuous features, and the rest are categorical.

\begin{table}[!htb]
    \centering
      \caption{Features from clinical/molecular data. The clinical/molecular data con- sists of 15 features with subcategories of the feature, including 3 continuous and 12 categorical features.}
    \label{tab:xgbFeatures}
    \begin{tabular}{l|p{12.5cm}}
    \hline
      \multicolumn{2}{c}{ \textbf{Categorical Features} with encoding} \\ \hline
     1&  Sex (1-Male, 0-Female) \\
     2&  Histology (0: Astrocytoma, 1:Glioblastoma, 2: Oligoastrocytoma, 3: Oligodendroglioma)  \\
     3& Grade (0:G2, 1:G3, 2:G4)  \\
     4&   MGMT promoter status (-1:NaN, 0:Methylated, 1:Unmethylated)   \\
      5&  Chr 7 gain/Chr 10 loss (-1:NaN, 0:Gain chr 7 \& loss chr 10,  1:No combined CNA)   \\
      6&  TERT promoter status (-1: NaN, 0: Mutant, 1: WT)  \\
        7& BRAF V600E status (-1: NaN, 0: WT )  \\
       8&  Transcriptome Subtype (-1: NaN, 0: CL, 2: NE, 1: ME, 3: PN)   \\
       9&  Pan-Glioma RNA Expression Cluster (-1: NaN, 0: LGr4, 1: unclassified)   \\
       10& Pan-Glioma DNA Methylation Cluster ( -1: NaN, 1: LGm5, 0: LGm4)   \\
      11&   Supervised DNA Methylation Cluster (-1: NaN, 0: Classic-like, 1: Mesenchymal-like )   \\
      12&  Random Forest Sturm Cluster ( -1: NaN, 0: Mesenchymal, 1: RTK II 'Classic',)    \\ \hline
       \multicolumn{2}{c}{ \textbf{Numerical Features}} \\ \hline
      13&  Age (years at diagnosis) \\
       14 &  Mutation Count \\
      15&   Percent Aneuploidy \\ \hline
    \end{tabular}
  
\end{table}


\subsection{Classification Threshold for OS}
After discussion with clinical experts and considering equal quartiles from the median OS, the boundary for short and long OS was set as 9 months and 13 months, respectively, to be of clinical significance/action. Cases with OS between these values were excluded to avoid confusion while dealing with instances near the threshold boundary. As a result, this study's short and long survivor classes having 94 cases each, hence avoiding class imbalance issues. The experimentation is performed in a 10-fold cross-validation Monte Carlo \cite{xu2001monte} configuration, while the data were divided randomly and proportionally in training (80\%), validation (10\%), and testing (10\%) datasets.




\begin{figure}[t]
    \centering
    \includegraphics[width=12.5cm]{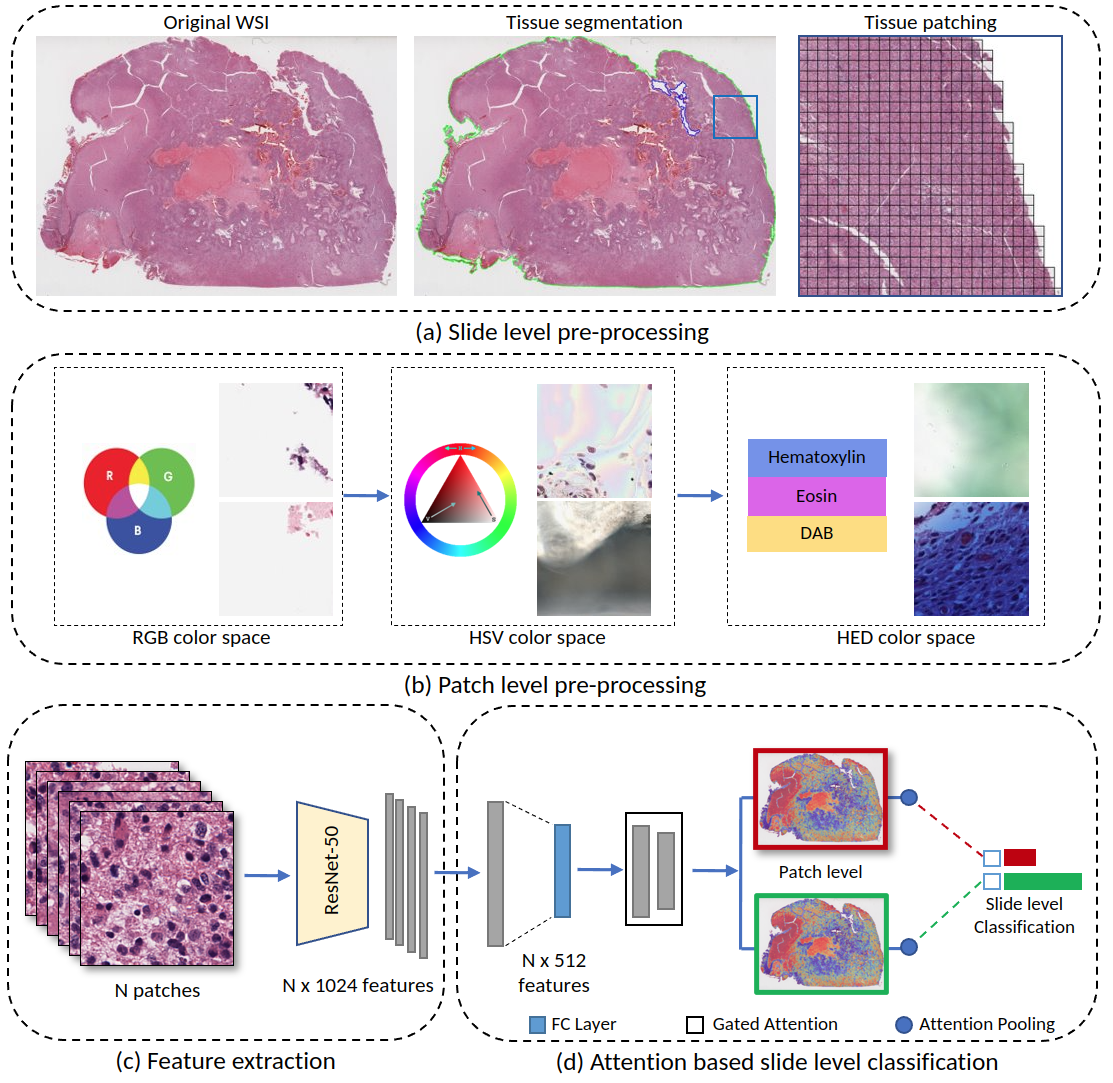}
    \caption{Schematic representation of the complete proposed methodological workflow.}
    \label{fig:blockdiagram}
\end{figure}

\subsection{Imaging Data Analysis}
The first step of the applied approach (as shown in Fig. \ref{fig:blockdiagram}(a)) is tissue segmentation, to distinguish the apparent tissue from the slide background. To facilitate this segmentation, the WSI is converted to the HSV color space in a downsampled resolution (second level of the WSI pyramid - i.e., downsampling by a factor of 16). A binary mask is created by thresholding the WSI's saturation channel, revealing the main tissue-occupied area. The mask is blurred using a median filter and then morphologically closed. This filtered tissue contour is used for further patch extraction by dividing it into rectangular patches of size $256\times 256$ at 20X magnification level. The coordinates of each patch, along with its corresponding slide metadata, are stored for the feature extraction process.

 
Further comprehensive patch-level image curation and tissue segmentation from WSI are deemed essential to differentiate between the tissue-occupied area and artifacts. This curation should exclude patches containing artifacts such as glass reflection, pen markings, black lines on the slide, and tissue tearing. The patch-level pre-processing involves three steps, as illustrated in Figure \ref{fig:blockdiagram}(b), with example patches that each step accounts for.
First, based on the intensity values in the Red-Green-Blue (RGB) space, patches within the obtained binary mask containing substantial white background (intensity $\ge$ 230) or black regions (intensity $\le$ 25) are discarded if the percentage of such pixels within the patch is greater than 60\%. However, this step alone cannot eliminate patches of glass reflection, dirt on the glass, or scanning artifacts.
To address such scenarios, patches are converted to the Hue-Saturation-Value (HSV) space, and the percentage of pixels in `S`(intensity $\le$ 25) and `V`(intensity $\ge$ 230) is checked. The patch is eliminated if this percentage exceeds the pre-defined threshold (95\%).
Finally, to remove patches with pen markings, the patches are converted into Hematoxylin-Eosin-DAB (HED) space by stain deconvolution \cite{ruifrok2001quantification}. Patches are eliminated if more than 80\% of pixels in the Eosin channel have an intensity $\le$ 50. After rigorous experimentation, all selected thresholds are determined empirically to ensure that only artifacts are removed and all tissue-occupied areas are preserved within the selected patches.

Fig. \ref{fig:blockdiagram}(c) shows feature extraction for each of the retained patches, leveraging a 1024-dimensional feature vector from a pre-trained ResNet-50 convolutional neural network. A multiple-instance-learning (MIL) framework with attention \cite{attention_mil} is then used to aggregate the patch-level extracted features into slide-level representations, and the final prediction as shown in Fig. \ref{fig:blockdiagram}(d). For each class (short- vs. long- survivor), the attention network ranks each region in the slide and assigns an `attention` score indicating its relative importance to the slide-level prediction. An additional binary clustering layer with 512 hidden neurons is introduced for learning class-specific features with SVM loss function after the first fully connected layer. The representative WSI patches with strong and weak attention scores train the clustering layer to learn a rich patch-level feature space, distinguishing between the positive and negative evidence of the two classes. The attention scores are also visualized as a heatmap to identify the important ROIs and contribute to appreciating the areas that drive the algorithmic predictions. A dense layer is used to aggregate high-importance patch predictions in the WSI (while ignoring those of low attention score) and classify each slide as belonging to either short- or long-survivors. Notably, the MIL-attention concept has been previously introduced as the clustering constrained attention multiple instance learning (CLAM) \cite{lu2021data}.

\subsection{Clinical Data Analysis} \label{sec:clinicalGBM_method}
XGBoost \cite{chen2016xgboost} is an optimized distributed gradient boosting library widely applied to machine learning tasks of structured data. We use the XGBoost model\footnote{\url{https://xgboost.readthedocs.io/en/stable/index.html}} to learn the classification task of detecting long vs. short survivors using clinical features listed in Table \ref{tab:xgbFeatures}.

To evaluate the performance of the XGBoost model on clinical features, we performed a stratified random sampling to create the training and the unseen test set comprising the 80\% and 20\% of the data, respectively. We use `gbtree' as the booster and set the objective as `binary:logistic' for the classification task. XGBoost model has the inbuilt capability to encode categorical variables and handle missing values. We have thus set parameters `enable\_categorical' to True and `missing' to Not a number (NaN). As a result, missing values (NaN) are encoded as -1 in Table \ref{tab:xgbFeatures}. XGBoost is an ensemble tree model; therefore, no feature scaling was performed. The train set was used to fine-tune hyperparameters using the 10-fold cross-validation approach. We set the parameters namely learning rate (eta)= 0.1, Minimum split loss (gamma) = 0.5,`max\_depth'=6, `subsample'= 0.6,`min\_child\_weight'=2, L2 regularization (lambda) = 1. 


\subsection{Late Fusion of Imaging \& Clinical Data}
Late fusion is performed to amalgamate the advantage of WSI patterns with clinical features. Specifically, we conduct the late fusion step by averaging the output probabilities obtained from the two distinct models at the decision level, using equal weights. The two distinct models mentioned here refer to the one model trained on the imaging features of the WSI data and the other trained on the clinical data (e.g., demographic information). The exact clinical data considered in this study are shown in Table~\ref{tab:xgbFeatures}. One of the primary advantages of late fusion is interpretability which allows the identification of decisive features for each sample in the dataset and can be used to evaluate the interactions between the features in each model.  
 
\subsection{Subgroup Analysis}
The subgrouping analysis intends to identify patient subgroups that may have different prognoses and responses to treatment. This approach can lead to more accurate predictions by accounting for the heterogeneity across the complete patient population. One of the methods of subgrouping analysis is to group patients based on clinical features that can be used to guide treatment decisions. Here we specifically seek prognostic stratification of GBM patients based on sex (male vs. female) and MGMT promoter methylation status (Methylated vs. Unmethylated). Using 10 folds generated using the Monte Carlo method\cite{xu2001monte}, we evaluate the performance of models trained using imaging and clinical data for subgroups. Table \ref{tab:subgroup_dataset} depicts the number of cases per subgroup identified for prognosis.
 
\begin{table}[!h]
    \centering
      \caption{Dataset for Subgrouping Analysis}
    \label{tab:subgroup_dataset}
\begin{tabular}{p{4cm}|cccc}
 
\hline
& Subgroup     & \#Samples & Short Survivors & Long Survivors \\ \hline
\multirow{2}{*}{Sex} &  Male         & 112           & 53              & 59             \\
& Female       & 76            & 41              & 35  \\ \hline

\multirow{2}{*}{MGMT Promoter Status } & Methylated   & 59            & 35              & 24             \\
& Unmethylated & 50            & 27              & 23             \\ \hline
\end{tabular}
\label{subgroup_dataset}
\end{table}
 
\section{Results}
\label{section:results}
 
 

\subsection{Imaging Results}
\label{sec:ImagingResults}

The quantitative performance evaluation, in terms of classification accuracy and the area under the curve (AUC) for each fold and the average across all folds is shown in Table \ref{tab1:accuracy} on the validation and test data.

\begin{table}[!ht]
\centering
\caption{Performance of prognostic stratification on validation and test data across various folds.}
\begin{tabular}{c|c |c|c|c}
    \multirow{2}{*}{Fold}  & \multicolumn{2}{c}{Validation} & \multicolumn{2}{c}{Test}  \\ 
        & AuC & Accuracy & AuC & Accuracy \\ \hline
0    & 0.691& 0.722& 0.654& 0.611\\
1    & 0.605& 0.611& 0.63& 0.667\\
2    &0.753& 0.722& 0.691& 0.611\\
3    &0.642& 0.611& 0.704& 0.778\\
4    & 0.63& 0.556& 0.519& 0.5\\
5    & 0.63& 0.5& 0.667& 0.611\\
6    & 0.741& 0.778& 0.617& 0.556\\
7    & 0.432& 0.5& 0.321& 0.444\\
8    & 0.642& 0.556& 0.716& 0.556\\
9    &0.6& 0.556& 0.531& 0.5\\ \hline
Mean & 0.637& 0.611& 0.605& 0.583\\ 
\end{tabular}
\label{tab1:accuracy}
\end{table}

\begin{figure}[!t]
     \centering
     \begin{subfigure}{0.3\textwidth}
         {\includegraphics[width=4cm]{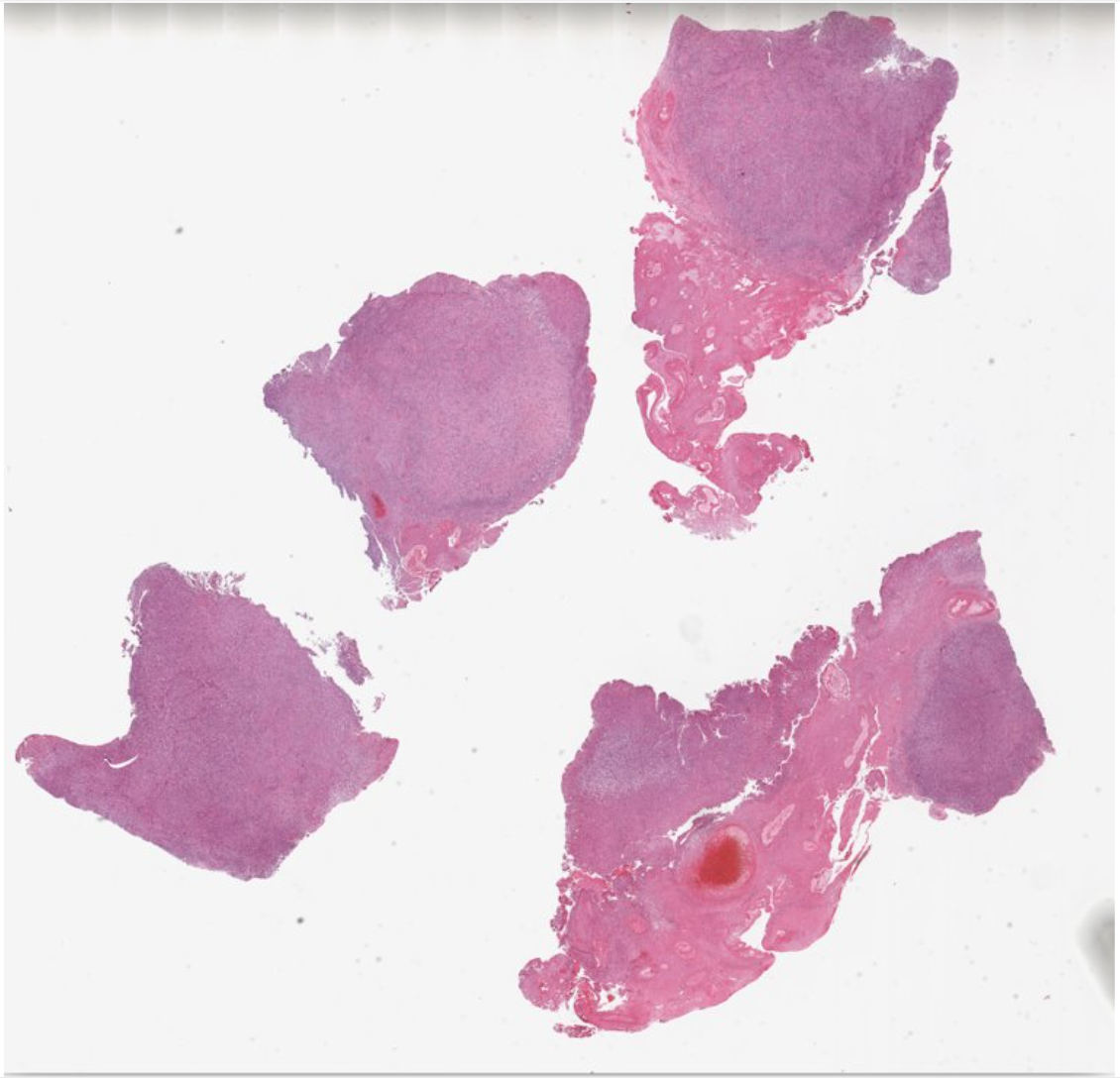}}
         \caption{Original WSI}
         \label{fig:sampleimage}
     \end{subfigure}
     \hspace{5mm}
     \begin{subfigure}{0.3\textwidth}
         {\includegraphics[width=4cm]{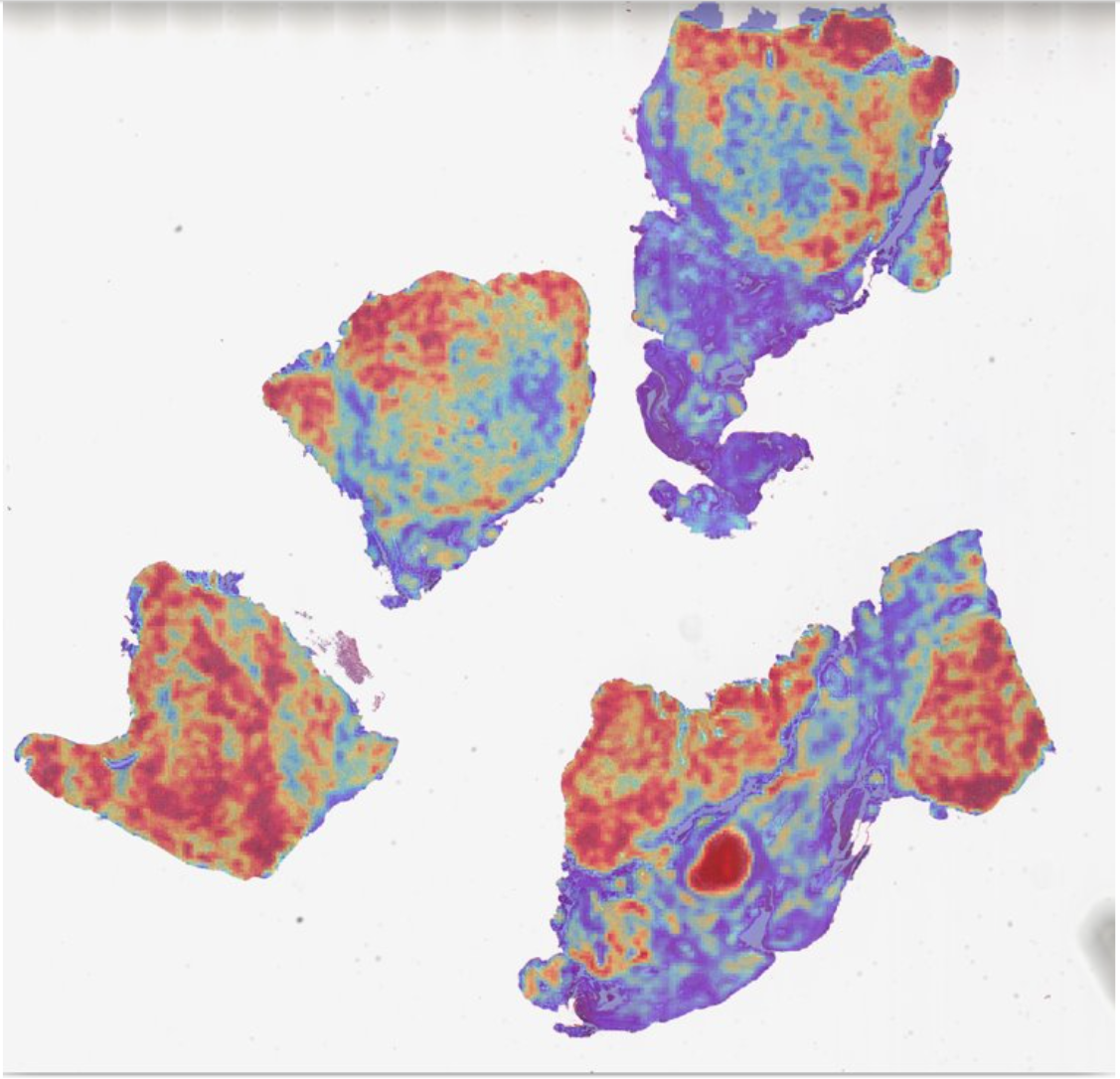}}
         \caption{Generated heatmap}
         \label{fig:sampleheatmap}
     \end{subfigure}
     \hfill
     \\
     \begin{subfigure}[b]{0.7\textwidth}
         {\includegraphics[width=9cm]{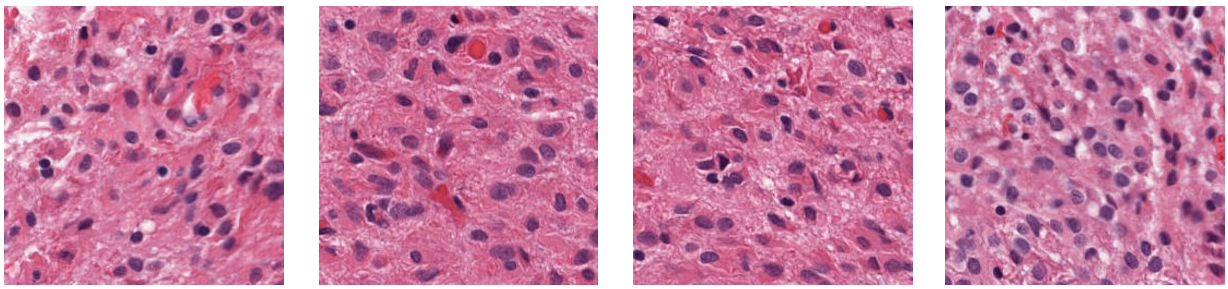}}
         \caption{Sample patches of high importance}
         \label{fig:samplepatches}
     \end{subfigure}
     \caption{Interpretability and visualization}
     \label{fig:interpretability}
\end{figure}         
Beyond the classification accuracy, this study focuses on the interpretability of the obtained predictions to understand what drives the particular algorithmic decision and attempt to further contribute to our clinical understanding of this disease.

Fig. \ref{fig:interpretability} shows an example WSI and its corresponding heatmap that aids in visualizing and interpreting each region's relative importance and morphological pattern in the WSI. It also shows some of the sample patches with high attention, i.e., identified as high importance for the final decision by the algorithm itself that can be useful for AI-assisted clinical diagnoses.

\subsection{Clinical Model Results} \label{sec:xgboost_clinical_results}

\begin{figure}
    \centering
    \includegraphics[width=\textwidth]{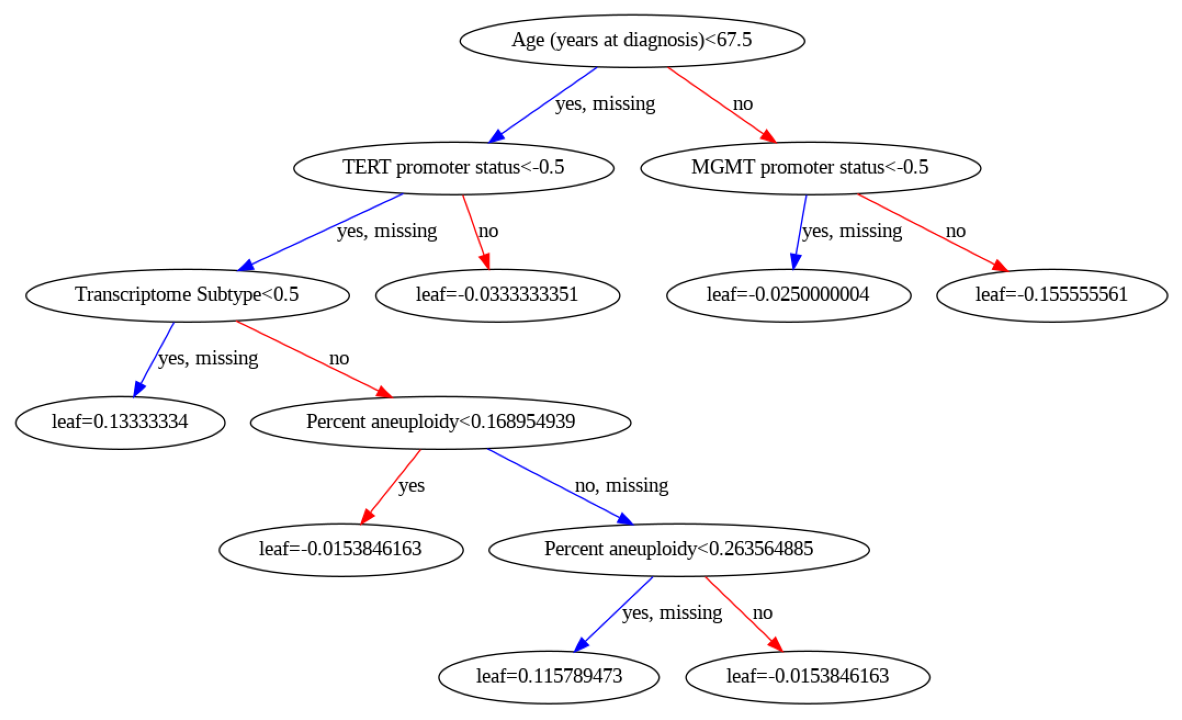}
    \caption{The First Decision Tree in trained XGBoost Model.  }
    \label{fig:xgb_tree}
\end{figure}

Using the learned hyperparameters, we trained an XGBoost model on the complete training set and evaluated its performance on the unseen testing set as created in Section \ref{sec:clinicalGBM_method}. The trained model provided an AUC of $0.622$ and an accuracy of $0.553$. We provide the first decision tree in the XGBoost model in Fig \ref{fig:xgb_tree}. The text in the node represents the split condition. For instance, the first node is `Age (years at diagnosis) $<67.5$'. The text on edge indicates the three cases: yes, that is, the split condition is true; no, that is, the split condition is False; and missing, which indicates the path if the variable is missing. The edge may lead to another node having a split condition or a leaf with a value indicating the log odds.

\begin{figure}
    \centering
    \includegraphics[width=\textwidth]{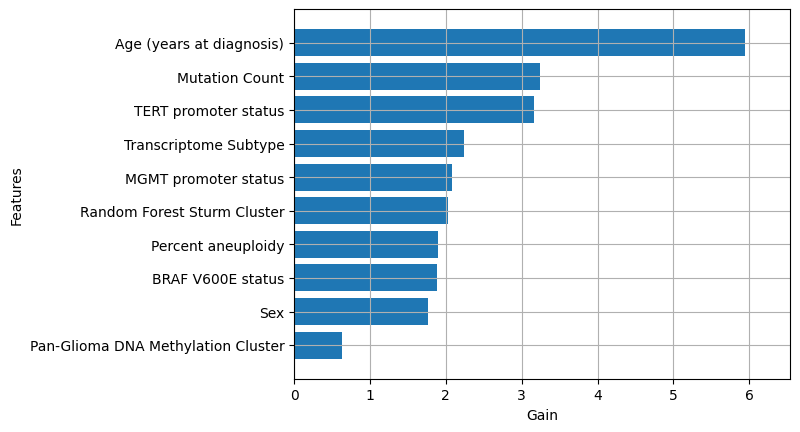}
    \caption{Feature Importance wrt gain that is, overall \textit{goodness} of a feature to determine the optimal split/branching across all trees. The higher the gain, the more significant the feature while making the prediction. Age has the highest gain followed by Mutation Count and TERT promoter status.}
    \label{fig:xgb_featureGain}
\end{figure}

In Fig. \ref{fig:xgb_featureGain}, we illustrate the top-10 features based on ``gain''. The gain indicates the overall \textit{goodness} of a feature used to determine the optimal split across all trees. Age (years at diagnosis) has a gain of $5.94$, making it the most significant feature followed by Mutation count ($3.24$). 

\subsubsection{SHAP}  
\begin{figure}
    \centering
    \includegraphics[scale=0.5]{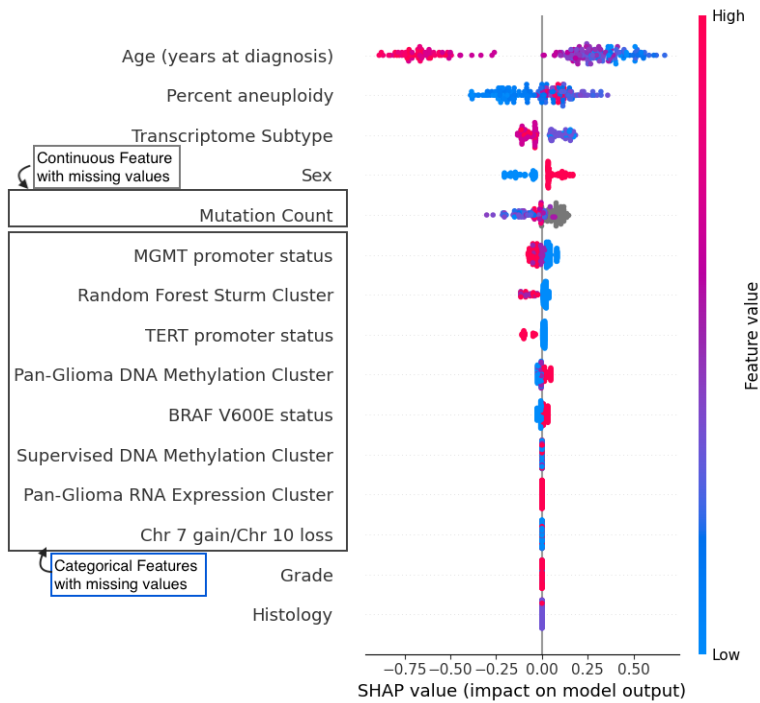}
    \caption{SHAP Summary Plot illustrating all 15 features  considered for training XGBoost. The features are arranged in decreasing order of impact on the model's output. Gray color indicates missing values for numerical features.  Sex, histology, and grade have no missing values. For the rest of the categorical features, the Blue color indicates missing values.}
    \label{fig:shap_summaryPlot}
\end{figure}

SHapley Additive exPlanations (SHAP) is a game theoretic approach widely used to interpret the effect of varying values of features on the model's prediction. For instance, SHAP values of a feature can be used to understand how increasing age affects the prediction of long OS when studying GBM prognosis. We depict the impact of each feature on the model's prediction in Fig \ref{fig:shap_summaryPlot}. It may be noted that the gray color for continuous features indicates missing values, as depicted for Mutation count. Sex, histology, and grade have no missing values (See Table \ref{tab:xgbFeatures}). For the rest of the categorical features, `Blue color' indicates missing values (since originally encoded as -1) and should be ignored.

\begin{figure}
     \centering
     \begin{subfigure}[b]{0.55\textwidth}
         \centering
         \includegraphics[width=\textwidth]{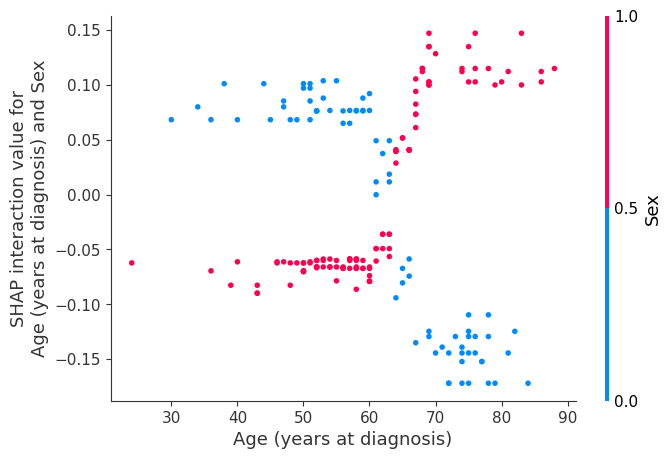}
         \caption{Interaction between Age and Sex in SHAP. Here, Red points represent Male (encoded as 1) and Blue points indicate Female samples (encoded as 0).}
         \label{fig:gender_interaction}
     \end{subfigure}
     \hfill
     \begin{subfigure}[b]{0.4\textwidth}
         \centering
         \includegraphics[width=\textwidth]{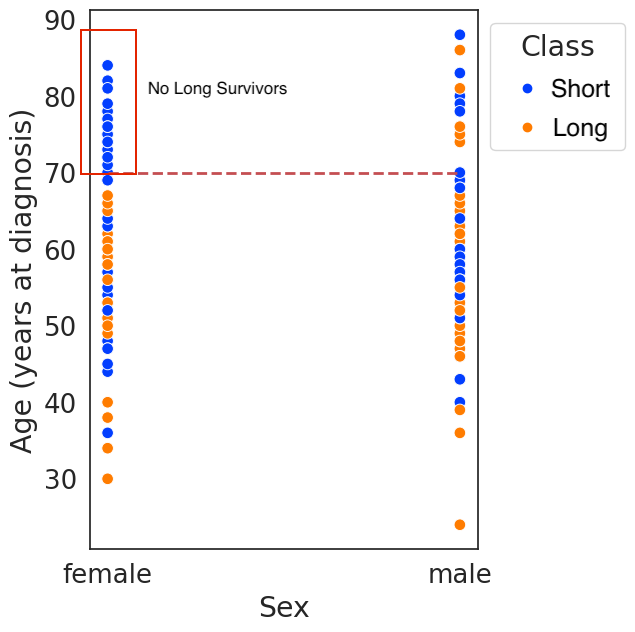}
         \caption{Distribution of Age and Sex in Original Data. There are no female long survivors over the age of 70. }
         \label{fig:original_genderDist}
     \end{subfigure}
        \caption{SHAP Interaction effect between Age and Sex on model outcome}
        \label{fig:gender_age_shap}
\end{figure}

\subsubsection{Comparison with Imaging Results}

Using the same 10-folds generated using Monte Carlo cross-validation \cite{xu2001monte} in Sec \ref{sec:ImagingResults}, we evaluate the quantitative performance of the XGBoost model trained on only clinical features. We use the same set of metrics: classification accuracy and the AUC for each fold. The results are provided in Table \ref{tab:xgb_clinicaldata}.

\begin{table}[!h]
    \centering
        \caption{Performance of XGBoost Model trained on clinical data for predicting Prognostic Stratification on Validation and Test data on folds generated using Monte Carlo Method.}
    \label{tab:xgb_clinicaldata}
    
    \begin{tabular}{c|c |c|c|c}
    \multirow{2}{*}{Fold}  & \multicolumn{2}{c}{Validation} & \multicolumn{2}{c}{Test}  \\ 
        & AuC & Accuracy & AuC & Accuracy \\ \hline
0&0.735&0.667&0.864&0.833 \\
1&0.481&0.500&0.691&0.611 \\
2&0.716&0.667&0.481&0.556 \\
3&0.494&0.556&0.790&0.722 \\
4&0.698&0.611&0.790&0.833 \\
5&0.753&0.667&0.481&0.556 \\
6&0.691&0.556&0.852&0.833 \\
7&0.691&0.667&0.728&0.556 \\
8&0.790&0.667&0.753&0.556 \\
9&0.623&0.611&0.716&0.667 \\ \hline
Mean &0.667&0.617&0.715&0.672
     
    \end{tabular}
    
\end{table}

\subsection{Late Fusion Results}
The probabilities returned by Imaging Model and XGBoost Model were averaged to perform a late fusion of both models. Based on these averaged probabilities, the quantitative performance evaluation, in terms of classification accuracy, the AUC for each fold, and the average across all folds, is shown in Table \ref{tab:fusionModel} on the validation and test data.
\begin{table}[!h]
    \centering
        \caption{Performance of fusion model for predicting prognostic stratification on Validation and Test data on folds generated using Monte Carlo Method.}
    \label{tab:fusionModel}
 \begin{tabular}{c|c |c|c|c}
    \multirow{2}{*}{Fold}  & \multicolumn{2}{c}{Validation} & \multicolumn{2}{c}{Test}  \\ 
        & AuC & Accuracy & AuC & Accuracy \\ \hline
 
0&0.778&0.667&0.864&0.722 \\
1&0.716 &0.722&0.704&0.611  \\
2&0.840&0.778&0.728&0.611  \\
3&0.617&0.611&0.642&0.722  \\
4&0.728&0.611&0.790&0.778  \\
5&0.840&0.722&0.691&0.611  \\
6&0.753&0.611&0.790&0.667  \\
7&0.790&0.667&0.691&0.611  \\
8&0.889&0.833&0.765&0.556  \\
9&0.778&0.667&0.790&0.667  \\ \hline
\textbf{Mean} &0.773&0.689&0.746&0.656  \\
    \end{tabular}
\end{table}

\subsection{Subgrouping Model Results}

\begin{table}[!htb]
    \centering
        \caption{Subgroup Results on 10 fold Cross-Validation}
    \label{tab:subgroup}
    \begin{tabular}{c|c|c|c|c|c|c|c}
   \multicolumn{2}{c|}{} & \multicolumn{2}{c|}{Imaging} & \multicolumn{2}{c|}{XGBoost} & \multicolumn{2}{c}{Fusion}\\ \hline
  &  SubGroup   &  AUC & Accuracy & AUC & Accuracy & AUC & Accuracy \\ \hline
    \multirow{2}{*}{Sex}   & Male (112) & 0.62 & 0.527 &0.652 & 0.6 & 0.697 & 0.627 \\ 
     &Female (76) &  0.725 & 0.638 & 0.712 & 0.662 & 0.8 & 0.725 \\ \hline
     \multirow{2}{*}{MGMT} & Methylated (50) & 0.6 & 0.683 & 0.694 & 0.683 & 0.713 &  0.683\\
      & Unmethylated (59)& 0.717 & 0.660 & 0.625 & 0.66 & 0.783 & 0.62 \\ \hline
    \end{tabular}

\end{table}
To identify the complex interaction among the subgroups of Sex and MGMT, we perform subgroup analysis by training models separately for Sex subgroups (that are, Male and Female) and MGMT subgroups (Methylated and Unmethylated) See Table \ref{tab:subgroup_dataset} for distribution of short and long survivors.


\section{Discussion}
    \label{section:discussion}

In this study, we presented an approach for prognostic stratification of patients diagnosed with GBM based on computational AI-based algorithms. The proposed workflow seamlessly integrates the WSI and clinical data for OS prediction. Our approach uses a Clustering-constrained attention-based Multiple Instance Learning (MIL) framework to classify WSI into short and long OS. By analyzing the attention heatmaps between short and long survivors, we can identify the vital histologic regions in the WSI that are significant for predicting OS. We also leverage clinical data to identify patient subgroups with distinct prognoses using the advanced XGBoost algorithm. Finally, we perform a late fusion of imaging features and clinical data to enhance the performance of our approach further.

OS prediction using both WSI and clinical data is an active area of research in medical imaging and machine learning. Combining these two data types can provide a more comprehensive view of a patient's condition and prognosis. To predict OS using WSI, a common approach is extracting features from WSI and feeding them as input. WSI can provide information on the histological features of a patient's tissue sample. Clinical data, including demographic information, can also be used to identify the heuristic characteristics of the patients. To effectively integrate both types of data, various approaches have been proposed. Separate Models can be trained on each data type, and their output can be combined in the late fusion step. 

In addition to the stratification accuracy, this study further focuses on detecting morphological patterns of prognostic relevance using a multiple-instance learning algorithm. The applied approach generates attention scores for individual WSI patches. It provides a visualization of a high-resolution heatmap for the complete WSI, identifying the important morphology patterns that drive the final WSI-/patient-level algorithmic prediction. Such visualization heatmap can contribute to the interpretability of the algorithmic decisions and significantly provide insights beyond, and contribute to, our current clinical knowledge of prognostically relevant histologic regions. Fig. \ref{fig:sampleimage} shows an example WSI with its corresponding heatmap.
 
\begin{figure}[!t]
     \centering
     \begin{subfigure}[b]{0.45\textwidth}
         {\includegraphics[width=5.8cm]{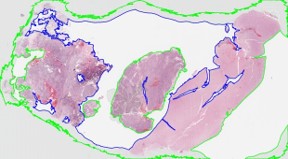}}
         \caption{WSI with artifacts }
         \label{fig:sampleimage}
     \end{subfigure}
     \hfill
     \begin{subfigure}[b]{0.45\textwidth}
         {\includegraphics[width=5.8cm]{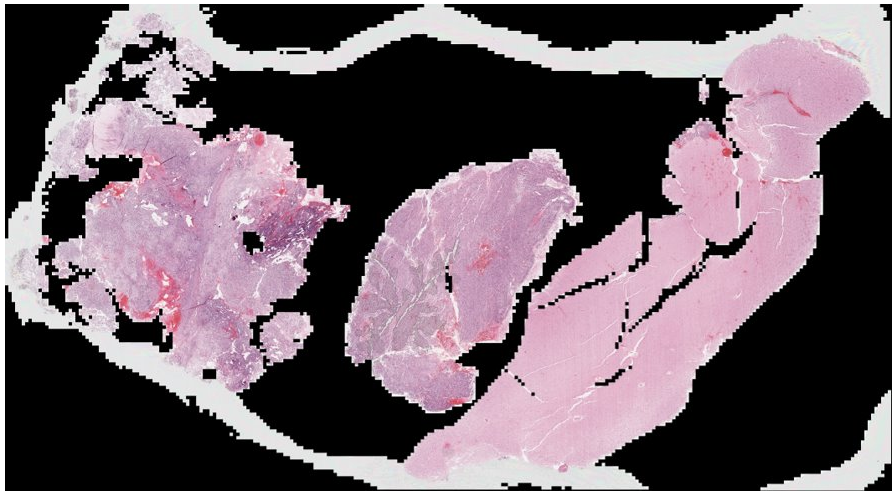}}
         \caption{Output of tissue segmentation}
         \label{fig:segimage}
     \end{subfigure}
     \begin{subfigure}[b]{0.45\textwidth}
         {\includegraphics[width=5.8cm]{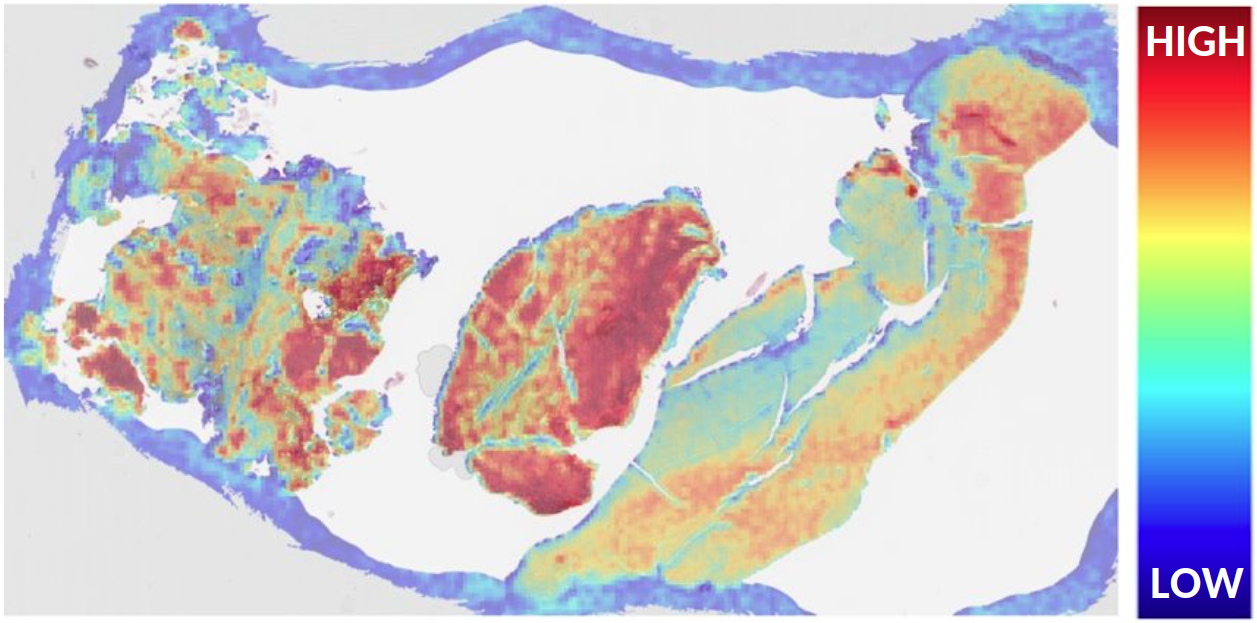}}
         \caption{Heatmap without pre-processing}
         \label{fig:sampleheatmap}
     \end{subfigure}
     \hfill
     \begin{subfigure}[b]{0.45\textwidth}
         {\includegraphics[width=5.8cm]{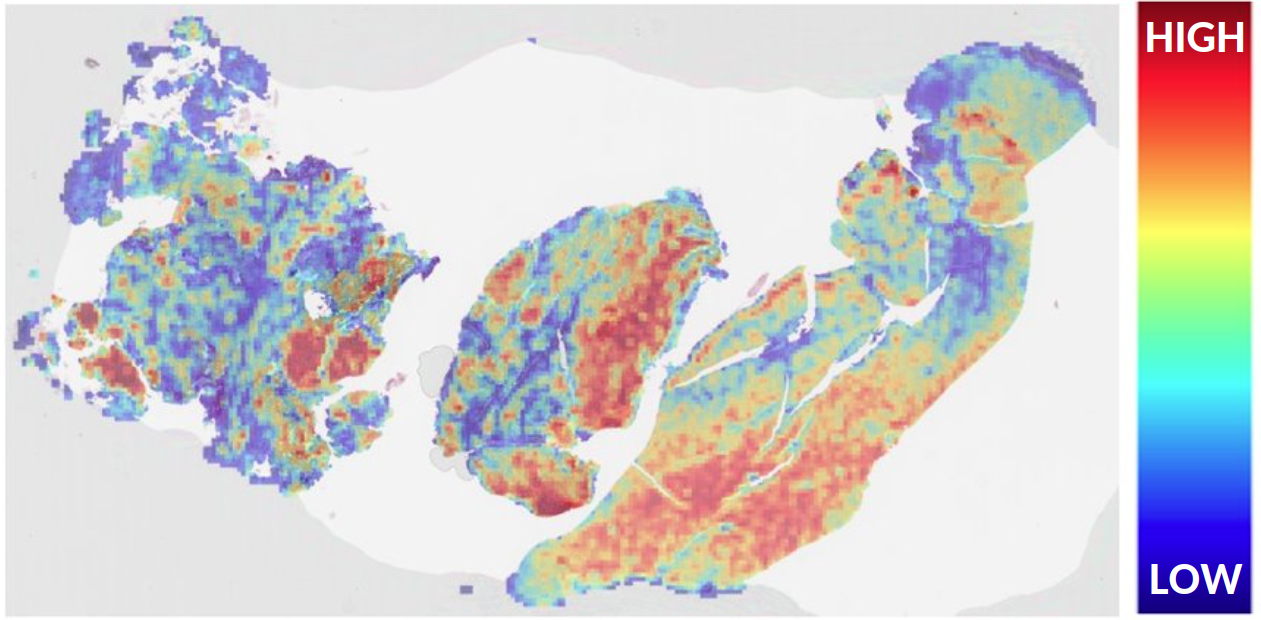}}
         \caption{Heatmap with pre-processing}
         \label{fig:newheatmap}
     \end{subfigure}
     \caption{Effect of pre-processing on the generated attention heatmap.}
     \label{fig:heatmap_preprocessing}
\end{figure}

Our approach for WSI is based on utilizing the entire WSI of each patient rather than extracting a subset of patches from a WSI (as typically done in the literature) or by randomly selecting a fixed number of patches from a WSI to obtain a global/patient-level prognostic decision \cite{mobadersany2018predicting, bejnordi2017diagnostic}. This should allow for capturing the heterogeneous morphology of GBM and ensure capturing all patterns of potential prognostic relevance.
Comprehensive image curation has proven essential for more optimal identification of morphology patterns and interpretation results. As explained in section \ref{section:methods}, the first step in our pipeline is tissue segmentation, and its output is overlayed in Fig. \ref{fig:sampleimage}. The region covered inside green boundaries is the identified tissue region by the algorithm for further processing, whereas the region to be excluded (non-tissue region) is bounded by blue boundary. The output of the original tissue segmentation algorithm is shown in Fig. \ref{fig:segimage}. As can be observed, the algorithm initially detected some artifactual background content as tissue, influencing the algorithm's decision. The algorithm has given low attention to this region, as shown in Fig. \ref{fig:sampleheatmap}, but ideally it should have been completely excluded from the analysis. It was observed in some of the cases that the interpretability algorithm gives great attention to pen markings, and that drives the decision, which is undesirable. To overcome this issue, we incorporated the additional pre-processing steps as illustrated in section \ref{section:methods} to eliminate various artifacts, such as glass reflections, dust, and pen markings, from the WSI.  After the additional pre-processing steps, the new heatmap is more accurate and focused, as shown in Fig. \ref{fig:newheatmap}.

The morphological patterns identified as of high diagnostic importance from the heatmap are discussed with neuropathologists (MPN) to understand if pathologists recognize them as clinically significant.  
Qualitative visualization of heatmaps contributes to the interpretability of the algorithmic decisions and provides insights into prognostically relevant histologic regions. For correctly predicted long survivors, the heatmaps indicate high attention throughout histologically malignant areas, including necrosis, hypercellularity, atypia, infiltration, and proliferation. For incorrectly predicted long-survivors, necrotic areas were not heavily weighted/attended, and we observed gemistocytic cells in densely packed. For some correctly predicted short-survivors, infiltration of gemistocytic cells without frank malignant features was present. Although the leptomeninges was generally not highly weighted on the heatmaps, foci within them containing concerning atypical cells potentially indicative of the leptomeningeal spread of tumor were highly weighted. 
After this qualitative investigation, we found that the attention heatmaps can often be reconciled with the pathologist's opinion, but in depth analysis is necessary for a deeper understanding. In addition, these heatmaps can also be used to analyze the misclassified cases and analyze their characteristics. The finding demonstrates that the proposed approach has the potential to be used for meaningful WSI-level interpretability and visualization in prognostic stratification for clinical research purposes and to identify morphological features associated with OS.  

The XGBoost model trained on clinical features provided a mean AUC of $0.67$ on the validation set and $0.715$ on the test set (See Table \ref{tab:xgb_clinicaldata}). XGBoost model superseded the model trained on Imaging features and showed significant gains in mean AUC and mean accuracy on both validation and test sets. This does reflect the significance of clinical features in predicting OS. Fig. \ref{fig:xgb_featureGain} illustrates ten features with non-zero gain. Age is the continuous feature with the highest gain, followed by Mutation count and TERT Promoter Status. It is worth noting that XGBoost performance on 10-fold Monte Carlo cross-validation (i.e., AUC= 0.715 and accuracy = 0.672 (See Table \ref{tab:xgb_clinicaldata}) ) is much higher than what we observed on the unseen test set (i.e., AUC=0.622 and accuracy=0.553) in Sec \ref{sec:xgboost_clinical_results}. 



For SHAP analysis, we only discuss Top-5 features (see Fig. \ref{fig:shap_summaryPlot}) having a significant impact on model output. As observed in XGBoost feature importance (See Fig. \ref{fig:xgb_featureGain}), age is the most significant feature. We can also note that high age negatively impacts the chances of long survival. Percent Aneuploidy has the second highest effect on model output after age, and the low value of Percent aneuploidy negatively impacts the chances for long survival. Sex is another demographic feature where the male sex positively affects the long-survivor class. We further studied the interaction effect between sex and age on the model's outcome (See Fig. \ref{fig:gender_interaction}) to understand how sex affects survival with increasing age. The female sex below the age of 70 positively influences the long survivor class, whereas the male sex has a slightly negative effect on Long survivor class prediction. However, the pattern switches drastically for patients above 70 years. Fig \ref{fig:original_genderDist} provides the distribution of samples with respect to age and sex in the data. It can be seen that there are no long female survivors above the age of 70, whereas long male survivors are relatively dispersed. Due to the lack of long-survival female patients, it seems that the SHAP values disproportionately assigned the positive effect of the male sex on the model's outcome to the long survival class. However, if we study patients below 70, female patients tend to have better chances for long survival than male patients. We also observed this effect in sub-group analysis for sex.

We began by evaluating our models using the WSI data. This dataset provided us with a large collection of images that were used to train and test our models. We then extended our evaluation to include clinical data. This additional modality allowed us to test our models' performance using different data types and validate our findings.  
The results of our study show that fusing the imaging and clinical data together significantly improved the performance of our models. Specifically, we found that our models were able to achieve better performance when the two modalities were combined, as compared to when they were used separately. In particular, we noticed that the fusion of imaging and clinical data was particularly effective in improving the performance of the models in cases where the performance of one modality alone was inferior (Tables \ref{tab1:accuracy} and \ref{tab:xgb_clinicaldata}). For example, in cases where the imaging modality had low performance, we were able to improve it through fusion with the clinical data, and vice versa. These results suggest that the integration of multiple modalities can be a powerful approach for improving the performance of machine learning models, particularly in cases where individual modalities are limited or suboptimal.

Analyzing the patient OS in these subgroups may provide insights into the differences in response to treatment and prognosis, which could lead to more personalized treatment options for patients with glioblastoma.

Sex-specific subgrouping is also evaluated in this study imaging, clinical and fusion evaluation. As seen from Table \ref{tab:subgroup}, the results on the female subgroup are consistently better than the metrics obtained for the male subgroup, irrespective of the model. The AUC goes up to 0.8 for the female subgroup, where it goes up to only 0.697 for the male subgroup, a gap of ~10 points. XGBoost models have improved performance (for both AUC and accuracy) on the Methylated subgroup compared to the Unmethylated subgroup. However, the Imaging model has a higher AUC on the Unmethylated subgroup but lower accuracy when compared to the Methylated subgroup. A similar trend is observed in results obtained using the Fusion model. It is important to note that subgroups present under MGMT Promoter Status have very few samples, and although trends in the data are noted, conclusions cannot be drawn.. In the future, we would like to evaluate this on more extensive datasets across different populations for concrete evidence. Nevertheless, the varying performance across subgroups suggests that there may be important differences in the performance of the model depending on the subgroup of interest and that it may be important to consider these differences when interpreting the model results.

Collectively the results presented in this study tend to suggest that sex-specific analyses can improve patient prognostication. Sex specific differences in cancer are an understudied area of research and more of these studies are needed to make progress towards clinical translation using computational approaches. 
In future works, sex can be considered as a stratification factor for predictive modeling. 



Although this study establishes some baseline performance, further developments are needed to solidify the algorithm’s robustness for prognostic stratification of GBM patients and to provide an interpretability mechanism to discover and analyze patterns along with clinical data and subgroups. We observe limitations that highlight the need for further investigation. The histologic appearance of many long-survivor cases would lead a neuropathologist to consider the tumors as aggressive. In fact, all of these tumors are aggressive. However, the ability of the algorithm to correctly predict longer survival despite these areas suggests that there may be other histologic features that are absent in these cases that are more important for predicting those tumors leading to the most abbreviated survival. The hot areas were hypercellular for some incorrectly predicted short-survivors; however, these were a minority of these specimens, which predominantly demonstrated a less cellular fibroblastic, spindled appearance.This suggests that considering the histology of the whole slide is crucial. 

There are several notable strengths of this study. Firstly, the accuracy of distinguishing short- and long-GBM survivors is different from other studies reported in the literature [2]. Secondly, the cohort used in this study, reclassified from the TCGA-GBM and TCGA- LGG data collections, represents cases acquired since the 1980s, according to the related metadata. This raises the question of whether the findings will be validated on a larger current external validation set. Building upon this question, to further confirm the algorithmic generalizability, evaluation of out- of-distribution data acquired from sites not represented in the training data is necessary. Finally, quantification of the exact correlation needs to be performed between the patient OS and the morphologic patterns across the distinct histologic tumor sub-regions to understand the morphology underlying tumor behavior.

The significance of the remarkably heterogeneous morphology in GBM has intrigued pathologists through time, and we propose to re-approach the relevance of these features with new technological tools. Identifying patterns from WSI and clinical data associated with patient prognosis within the current integrated histologic-molecular definition of the GBM entity will allow the clinical neuropathologist to provide additional prognostic information gleaned during the microscopic assessment to the treating team as well as suggest avenues of biological investigation for understanding and potentially treating GBM. Also, this study's findings indicate that the identification of sex and MGMT subgroups among survivals can be considered as a factor for stratification among GBM patients. The successful development of such an algorithm can have a direct influence on patient management and decision-making for treatment selection, and may further contribute to our current biological understanding of GBM.


\bibliographystyle{splncs04}
\bibliography{main.bib}

\end{document}